\documentclass[twocolumn,showpacs,showkeys,preprintnumbers,amsmath,amssymb]{revtex4}
\usepackage [dvips]{graphicx}
\usepackage [english]{babel}

\begin{document}

\title{Energetics of metal slabs and clusters: the rectangle-box model}

\author{V.V. Pogosov\footnote{Corresponding author:
vpogosov@zntu.edu.ua}, V.P. Kurbatsky, E.V. Vasyutin}

\address{Department of Microelectronics and Semiconductor Devices,
Zaporozhye National Technical University, Zhukovsky Str. 63,
Zaporozhye 69064, Ukraine}

\date{\today}

\begin{abstract}
An expansion of energy characteristics of wide thin slab of
thickness $L$ in power of $1/L$ is constructed using the
free-electron approximation and the model of a potential well of
finite depth. Accuracy of results in each order of the expansion
is analyzed. Size dependences of the work function and electronic
elastic force for Au and Na slabs are calculated. It is concluded
that the work function of low-dimensional metal structure is
always smaller that of semi-infinite metal sample.

A mechanism for the Coulomb instability of charged metal clusters,
different from Rayleigh's one, is discussed. The two-component
model of a metallic cluster yields the different critical sizes
depending on a kind of charging particles (electrons or ions). For
the cuboid clusters, the electronic spectrum quantization is taken
into account. The calculated critical sizes of Ag$_{\rm N}^{\rm
2-}$ and Au$_{\rm N}^{\rm 3-}$ clusters are in a good agreement
with experimental data. A qualitative explanation is suggested for
the Coulomb explosion of positively charged Na$_{\rm N}^{\rm n+}$
clusters at 3$\leq n\leq$5.
\end{abstract}

\maketitle

\section{Introduction}

Clusters constitute a bridge  between atomic, molecular and
surface physics. Numerous investigations of physical properties of
the low-dimensional systems are stipulated by their promising
application in the nanotechnology.

Recently  \cite{14.},  the work function of atomically uniform Ag
films grown on Fe(100) was measured as a function of film thickness.
The maxima of the work function magnitude correspond to the
``magic'' thicknesses. Scanning-tunneling microscopy observation
of Pb nanocrystals grown on Cu(111) indicates that in the
equilibrium distribution of the island heights, some heights
appear much more frequently than other ones \cite{Otero}. An
appearance of these `magic'' island heights on the Cu flat surface
was studied by self-consistent electronic structure calculations
\cite{Chulok}.

Point contacts of gold bodies are investigated experimentally in
Refs. \cite{6.,7.} during elongation of the contacts to the
rupture. It is shown that oscillations of elastic constants appear
simultaneously with an abrupt change  in conductance. A
dimensionality of a contact varies during the process of
stretching. Indeed, at the moment of formation of a contact, the
contact region can be represented as a slab inserted between
electrodes, whereas at the moment of its rupture, the contact
region becomes a wire. Thus, in the experiment we observe a
transition from $2D$ (or $0D$) to  $1D$ open electron system.

Analytical approaches to the determination of the density of
states and the Fermi energy for metal slabs are proposed in Refs.
\cite{1.,Nag} on the basis of the free electron model.
Quantization of a contact potential difference for a slab was
described in Ref. \cite{2.} within the framework of the model of
hard walls. Jumps of the tensile force in a point contact were
explained in Refs. \cite{3.,4.,5.}. However, the work function
cannot be determined in this simplest model (see review
\cite{PR-2003}).

Probably,  the electron work function of a slab was  calculated
for the first time by Schulte \cite{Schulte}. The work function
magnitude showed oscillations near its average value. In Ref.
\cite{Pash}, the  above approaches have been criticised. Detailed
computations \cite{8.,9.,10.,Woi,11.,12.,13.} (including  \emph{ab
initio} calculations) performed to date do not yield an
unequivocal conclusion about size dependence of the work function
of isolated slabs and wires. In addition, amplitudes of the work
function oscillations are larger than in experiment.

Since the work of Sattler \emph{et al.} \cite{Sattler},
mass-spectrometric investigations of the charging effects in
cluster beams have clearly demonstrated the size-dependent Coulomb
instability of clusters composed of a countable number of atoms
\cite{Nah-1,Nah-2,Yannou}.

Rayleigh`s theory predicts an instability of a charged liquid
sphere of $R$ radius, when the Hartree  energy exceeds twice the
surface energy. The critical charge is defined by the expression
\begin{equation}
Q_{\rm R} = \pm\sqrt{16\pi R^{\rm 3}\tau}, \label{Rel}
\end{equation}
where $\tau$ is the surface tension (or stress). Recently, this
criterium  of stability has been confirmed  in the experiment
\cite{Duft} for microdroplets of ethylene glycol.

Eq. (\ref{Rel}) (i.e. Rayleigh`s criterium) does not determine,
which type of the particles charge the cluster. A metallic droplet
can contain either an excess number of electrons $\Delta N_{\rm
R}^{\rm e}=|Q_{\rm R}|/e$ or ions $\Delta N_{\rm R}^{\rm
i}=|Q_{\rm R}|/Ze$, where $Z$ is the valence and $e$ is the
elementary positive charge. Therefore, such a problem should be
considered using the two-component model of a cluster in which
electrons and ions are interpreted on equal footing
\cite{93,1996}. A sign of the excess charge results in different
size dependence,  $\Delta N_{\rm e,i}\propto R$ or $\Delta N_{\rm
R}\propto R^{\rm 3/2}$.

In the present work, we developed an analytical theory of
size-dependent energy and force characteristics for metal slabs
using an elementary one-particle approach. This simple model makes
it possible to calculate the oscillatory size dependence of the
work function and elastic force. Thermal effects are not
considered. An assumption of the ideal plastic strain (when volume
of the slab remains constant during stretching) allows for the
comparison of theoretical results and experimental ones \cite{6.}.

In the framework of the two-component model, the size-dependent
Coulomb instability of charged metallic clusters was described. A
spectrum quantization  is taken into account for a cluster having
the shape of parallelepiped. The model makes it possible to study
a physical origin of the instability.  Theoretical results on
critical sizes of different shaped charged Au, Ag and Na clusters
are in agreement with experimental ones (see Refs.
\cite{Nah-1,Yannou} and referencee therein).

\section{SLABS: QUASICLASSICAL APPROXIMATION}

\subsection{Formulation of problem}

We consider a thin slab with the thickness, $L_{\rm z}\simeq
\lambda_{\rm F0}$ ($\lambda_{\rm F0}$ is the Fermi wavelength of
electrons in a  semi-infinite metal). The thickness is much
smaller than other dimensions, $L_{\rm z}\ll L_{\rm x},L_{\rm y}$.
Thus, the discreteness of the electron momentum components $p_{\rm
x}$ and $p_{\rm y}$ can be ignored. For typical values of the
electron concentration in metals, we have $\lambda_{\rm F0}\simeq
$ 0.5 nm.

As a first approximation, the profile of the one-electron
effective potential in the slab can be represented as a
rectangular potential well of constant depth $U_{\rm 0}<0$ with
dimensions $L_{\rm x}$ , $L_{\rm y}$, and $L_{\rm z}$. A solution
of the three-dimensional Schr\"{o}dinger equation for a quantum
box is simple. An equation can be decoupled into three
one-dimensional equations. Therefore it is characterized by a set of
electron wave numbers $ k_{\rm xj}=2\pi j/L_{\rm x}$, $k_{\rm
ys}=2\pi s/L_{\rm y}$, and $k_{\rm zi}$, which are roots of the
equation
\begin{equation} k_{\rm zi}L_{\rm z}=-2\arcsin
(k_{\rm zi}/k_{\rm 0}) +\pi i, \label{1s}
\end{equation}
where $k_{\rm 0}=\sqrt{2m|U_{\rm 0}|}$ and $m$ is the electron
mass. The numbers $j, s = 0,\pm 1, \pm2, \pm3, ...$ and $i = 1, 2,
3, ...$

The set of wave numbers determines the electron energy:
$$
E_{\rm p}=\frac{\hbar^{\rm 2}}{2m}\left(k_{\rm xj}^{\rm 2} +k_{\rm
ys}^{\rm 2}+k_{\rm zi}^{\rm 2}\right),
$$
where $p$ is the number of the electron state (the states are
numbered in order of increasing of the electron energy). For a
cuboid with hard walls and dimensions $L_{\rm x}$, $L_{\rm y}$,
$L_{\rm z}$, we use the well-known expression
$$
E_{\rm p}^{\rm \infty}=\frac{\hbar^{\rm 2}\pi^{\rm 2}}{2m}
\left(\frac{j^{\rm 2}}{L_{\rm x}^2}+\frac{s^{\rm 2}}{L_{\rm y}^2}+
\frac{i^{\rm 2}}{L_{\rm z}^2}\right),
$$
where  $j,s,i$ are the natural numbers.

It is convenient to use dimensionless variables by choosing
$U_{\rm 0}$ as a unit of energy and $k_{\rm 0}^{\rm -1}$ as a unit
of length. We introduce the following notation:
$$
  \xi_{\rm xj}=k_{\rm xj}/k_{\rm 0}, \quad \xi_{\rm ys}=
  k_{\rm ys}/k_{\rm 0}, \quad \xi_{\rm i}=k_{\rm zi}/k_{\rm 0},
$$
$$
l_{\rm x}=k_{\rm 0}L_{\rm x}/2\pi, \quad l_{\rm y}=k_{\rm 0}L_{\rm
y}/2\pi, \quad l=k_{\rm 0}L_{\rm z}/\pi.
$$
Note that energy can be interpreted as the square of the state
vector in the $\xi-$space, $\xi_{\rm p}^{\rm 2}=\xi_{\rm xj}^{\rm
2} +\xi_{\rm ys}^{\rm 2}+\xi_{\rm i}^{\rm 2}$, with $\xi_{\rm
p}\leq 1$. Eq. (\ref{1s}) gets a form
\begin{equation}
l\xi_{\rm i}=-\frac{2}{\pi}\arcsin \xi_{\rm i} +i. \label{3s}
\end{equation}
Not only solutions of Eq. (\ref{3s}) but also their number  are
fully determined by the thickness $l$, namely, $i_{\rm F}=[l]+1$,
where $[a]$ denotes the integer part of $a$.

\subsection{Density of states}

Let us estimate the interval $\Delta \xi$ between neighboring
values of $\xi_{\rm z}$.  It follows from Eq. (\ref{3s}) that, for
sufficiently large values of $l$,
\begin{equation}
\Delta \xi\approx 1/l. \label{4s}
\end{equation}
Distances between two consecutive values of $\xi_{\rm x}$ and
$\xi_{\rm y}$ are small, namely, $\Delta \xi_{\rm x}=\xi_{\rm
xj+1}-\xi_{\rm xj}=1/l_{\rm x}$ and $\Delta \xi_{\rm y}=1/l_{\rm
y}$. For the relative values of the slab dimensions assumed by us,
it can be easily found that $\Delta \xi\gg \Delta \xi_{\rm
x},\Delta \xi_{\rm y}$. We see that the  electron states
$\left\{\xi_{\rm xj},\xi_{\rm ys},\xi_{\rm i}\right\}$ form a
system of parallel planes $\xi_{\rm z}=\xi_{\rm i}$ in $\xi-$space
and that the density of states on all these planes is the same and
equal to
\begin{equation}
\sigma =2/(\Delta \xi_{\rm x}\Delta \xi_{\rm y})= 2l_{\rm x}l_{\rm
y}. \label{5s}
\end{equation}
The factor $2$ takes into account two possible values of the
electron spin polarization.

Electrons occupy the states, beginning from the point
$\left\{0,0,\xi_{\rm 1}\right\}$, in ascending order of $\xi_{\rm
p}^{\rm 2}$, i.e., of the state energy. Therefore, it appears that
all occupied states lie in the $\xi$-space domain bounded by the
plane $\xi_{\rm z}=\xi_{\rm 1}$ and the hemisphere of radius
$\xi_{\rm F}=\sqrt{E_{\rm F}/|U_{\rm 0}|}$, where $E_{\rm F}>0$ is
the Fermi energy equal to the maximum energy of occupied states.

The occupied states are distributed with density $\sigma$  over
 disks formed by intersections of the Fermi hemisphere
with the planes $\xi_{\rm z}=\xi_{\rm i}$,  $i  = 1, 2,...,i_{\rm
F}$. The area of the disk is $S_{\rm i}=\pi(\xi_{\rm F}^{\rm
2}-\xi_{\rm i}^{\rm 2})$. The number of occupied states coincides
with the number of free electrons in the slab,
\begin{equation}
N_{\rm e}=\sigma\sum\limits_{\rm i=1}^{\rm i_{\rm F}}S_{\rm i}
=2l_{\rm x}l_{\rm y}\sum\limits_{\rm i=1}^{\rm i_{\rm
F}}\pi(\xi_{\rm F}^{\rm 2}-\xi_{\rm i}^{\rm 2}), \label{6s}
\end{equation}
where $i_{\rm F}$ is the number of roots of Eq. \ref{3s}.

The number of occupied states per unit volume is
\begin{equation}
\nu\equiv \frac{N_{\rm e}}{l_{\rm x}l_{\rm y}l}=
\frac{2\pi}{l}\left(i_{\rm F}\varepsilon_{\rm F}-\sum\limits_{\rm
i=1}^{\rm i_{\rm F}}\xi_{\rm i}^{\rm 2}\right), \label{7s}
\end{equation}
where we used the notation $\varepsilon_{\rm F}\equiv E_{\rm
F}/|U_{\rm 0}|= \xi_{\rm F}^{\rm 2}$.

By definition, the density of states $\rho(E)$ is the number of
states per unit energy interval near the energy $E$  and per unit
volume of the metal. In order to find this quantity, we write Eq.
(\ref{7s}) in the form
\begin{equation}
\nu= \frac{2\pi}{l}\left(i_{\rm \varepsilon}\varepsilon-
\sum\limits_{\rm i=1}^{\rm i_{\rm \varepsilon}}\xi_{\rm i}^{\rm
2}\right). \label{8s}
\end{equation}
One can interpret $\nu$ as a number of states (per unit volume)
whose energies do not exceed $\varepsilon$. In Eq. (\ref{8s})
$i_{\rm \varepsilon}$ is the index of the greatest of the roots of
Eq. (\ref{3s}) satisfying the condition  $\xi_{\rm i}^{\rm 2}\leq
\varepsilon$.

We find from Eq. (\ref{3s}) that
$$
i=l\xi_{\rm i}+\frac{2}{\pi}\arcsin \xi_{\rm i}.
$$
Substituting here  $\xi_{\rm i}$  by $\xi$, we let $\xi$ to take any
value in the limits from $\xi_{\rm 1}$ to 1 and form an
integer-valued increasing function $i(\xi)$ such that at the
points $\xi=\xi_{\rm i}$ the value of the function is increased by
one and in the intervals between these points the function does
not change. Substituting  $\xi=\sqrt{\varepsilon}$, we obtain
\begin{equation}
i_{\rm \varepsilon}=\left[l\sqrt{\varepsilon}+\frac{2}{\pi}
\arcsin \sqrt{\varepsilon}\right]. \label{9s}
\end{equation}
Square brackets indicate an integer part.

By differentiating l.h.s. of Eq. (\ref{8s}) with respect to $\xi$
under the condition $i_{\rm \varepsilon}=const$, we find
$\rho(\varepsilon)$. Using Eq. (\ref{9s}) and working backward
through the normalizations, we obtain
\begin{multline}
\rho(E)\equiv \frac{1}{V}\frac{dN_{\rm e}}{dE}\\=
\frac{m}{\pi\hbar ^{\rm 2}L_{\rm z}}\left[\frac{L_{\rm
z}\sqrt{2mE}}{\pi\hbar}+ \frac{2}{\pi}
\arcsin{\sqrt{\frac{E}{|U_{\rm 0}|}}}\right],
 \label{11s}
\end{multline}
where $V=L_{\rm x}L_{\rm y}L_{\rm z}$.

\subsection{Size dependence of the work function}

From Eq. (\ref{7s}), we obtain
\begin{equation}
\varepsilon_{\rm F}= \frac{1}{i_{\rm F}}\left(\frac{\nu l}{2\pi}
+\sum_{\rm i=1}^{\rm i_{\rm F}}\xi_{\rm i}^{\rm 2}\right).
 \label{12s}
\end{equation}
Using Eq. (\ref{9s}), we find that
\begin{equation}
i_{\rm F}=\left[l\sqrt{\varepsilon_{\rm F}}+\frac{2}{\pi}\arcsin
\sqrt{\varepsilon_{\rm F}}\right].
 \label{13s}
\end{equation}

In what follows, we assume that the electron density in the slab
does not depend on its size and is
\begin{equation}
\frac{N_{\rm e}}{V}=\frac{k_{\rm 0}^{\rm 3}}{4\pi^{\rm 3}}\nu
=\bar{n}_{\rm e}.
 \label{14s}
\end{equation}

If the depth of the well is fixed, then Eq. (\ref{14s}) implies
that $\nu=const$. The thickness dependence of the Fermi energy
$\varepsilon_{\rm F}(l)$ can be found by solving the set of
equations (\ref{12s}) and (\ref{13s}) under the additional
condition $\nu=const$.

Substituting the expression $\bar{n}_{\rm e}\equiv k_{\rm F0}^{\rm
3}/3\pi^{\rm 2}$ into Eq. (\ref{14s}) ($k_{\rm F0}$ is the Fermi
wave number for the semi-infinite metal), we find
\begin{equation}
\nu=\frac{4}{3}\pi\xi_{\rm F0}^{\rm 3}, \label{16s}
\end{equation}
where $\xi_{\rm F0}\equiv k_{\rm F0}/k_{\rm 0}$, i.e. $\nu$ is
equal to  doubled  volume of the Fermi hemisphere in $\xi-$space
in the limiting case $l\rightarrow \infty$.

Setting  $i_{\rm F}=const$, from Eq. (\ref{12s}) we obtain
\begin{equation}
\frac{d\varepsilon_{\rm F}}{dl}=\frac{1}{i_{\rm
F}}\left(\frac{\nu}{2\pi}+\frac{d}{dl} \sum_{\rm i=1}^{\rm i_{\rm
F}}\xi_{\rm i}^{\rm 2}\right), \label{19s}
\end{equation}
In order to see how the roots of Eq. (\ref{3s}) change with
varying $l$, we differentiate both parts of this equation and find
that
\begin{equation}
\frac{d}{dl}\xi_{\rm i}^{\rm 2}=-\frac{2\xi_{\rm i}^{\rm 2}}{l+
\frac{2}{\pi\sqrt{1-\xi_{\rm i}^{\rm 2}}}}\leq 0. \label{20s}
\end{equation}
Here, the equality valid only in the limit $l\rightarrow\infty$,
i.e., as  $\xi_{\rm i}\rightarrow 0$ for all $i$.

The disks  $\xi_{\rm z}=\xi_{\rm i}$ are lowered with increasing
$l$. The lowering rate decreases gradually, so that the lower
disks move more slowly that the higher ones. Accordingly, the
distance between the disks decreases and their number  $i_{\rm F}$
grows. It is seen from Eq. (\ref{13s}) that this number increases
by 1 each time the equality $\varepsilon_{\rm F}=\xi^{\rm 2}_{\rm
i_{\rm F}+1}$ is satisfied. The process of disk lowering is
accompanied by the ``pulsation'' of the Fermi hemisphere. Its
radius $\xi_{\rm F}=\sqrt{\varepsilon_{\rm F}}$ alternately
increases (as $d\varepsilon_{\rm F}/dl>0$), and decreases ( as
$d\varepsilon_{\rm F}/dl<0$), having the average tendency to
decrease. At $\xi_{\rm F}=\xi_{\rm i_{\rm F}+1}$ the derivative
$d\varepsilon_{\rm F}/dl$ is discontinuous. The value of the jump
decreases with increasing $l$.

The minimum value of $L_{\rm z}$ corresponds to the thickness
equal to the atom diameter. Let us estimate the minimum value of
$l$. We use $l=L_{\rm z}\sqrt{2m|U_{\rm 0}|}/(\pi\hbar)$, where
$L_{\rm z}$ = 0.5 nm (it means that we have a single layer of
atoms), and
\begin{equation}
|U_{\rm 0}|=E_{\rm F0}+W_{\rm e0}, \quad E_{\rm
F0}=\frac{\hbar^{2}}{2m}(3\pi^{2}\bar{n}_{\rm e})^{2/3}.
\label{Us}
\end{equation}
The work function for a semi-indefinite metal $W_{\rm 0}$ equals
to 2.25 and 4.25 eV for Cs and Al, respectively \cite{Mich}. As a
result, we have $1.6<l_{\rm min}<3.5$. We assume henceforth the
value $1/l$ to be small and apply an expansion in terms of $1/l$
for calculating of the Fermi energy.

Now we introduce the notation $\alpha\equiv 1/l$. The $\xi_{\rm
i}(\alpha)$ dependence is implicitly determined by Eq. (\ref{3s}),
which can be written as
\begin{equation}
\frac{\xi_{\rm i}}{\alpha}=-\frac{2}{\pi}\arcsin \xi_{\rm i}+i.
 \label{24s}
\end{equation}
We look for the roots   $\xi_{\rm i}$ in the form of the expansion
\begin{equation}
\xi_{\rm i}=\xi_{\rm i}\left|_{_{\rm \alpha=0}}\right.+\xi_{\rm
i}^{\rm \prime}\left|_{_{\rm \alpha=0}}\right.\alpha+
\frac{1}{2}\xi_{\rm i}^{\rm \prime\prime}\left|_{_{\rm
\alpha=0}}\right.\alpha^{\rm 2}+ \frac{1}{6}\xi_{\rm i}^{\rm
\prime\prime\prime}\left|_{_{\rm \alpha=0}}\right.\alpha^{\rm
3}+...
 \label{25s}
\end{equation}
Keeping the terms to the order of $\alpha^{\rm 3}$, we obtain the
Fermi energy up to the order  $\alpha^{\rm 2}$. It is seen from
Eq. (\ref{24s}) that
\begin{equation}
\xi_{\rm i}\left|_{_{\rm \alpha=0}}\right.=0, \quad \xi_{\rm
i}/\alpha \left|_{_{\rm \alpha=0}}\right.=i.
 \label{26s}
\end{equation}
Differentiating both sides of Eq. (\ref{24s}) with respect to
$\alpha$ and multiplying the result by $\alpha$, we obtain
\begin{equation}
\xi_{\rm i}^{\rm \prime}-\frac{\xi_{\rm i}}{\alpha}=-\frac{2}{\pi}
\frac{1}{\sqrt{1-\xi_{\rm i}^{\rm 2}}}\alpha \xi_{\rm i}^{\rm
\prime}.
 \label{27s}
\end{equation}
Setting  $\alpha=0$, we find
\begin{equation}
\xi_{\rm i}^{\rm \prime}\left|_{_{\rm \alpha=0}}\right.=i.
\label{28s}
\end{equation}
In a similar way, we obtain
$$
\xi_{\rm i}^{\rm \prime\prime}\left|_{_{\rm
\alpha=0}}\right.=-\frac{4i}{\pi},\quad \xi_{\rm i}^{\rm
\prime\prime\prime}\left|_{_{\rm
\alpha=0}}\right.=-\frac{24i}{\pi^{\rm 2}}.
$$
Substituting the obtained expressions into Eq. (\ref{25s}), we
find
\begin{equation}
\xi_{\rm i}= i\alpha-\frac{2i}{\pi}\alpha^{\rm
2}+\frac{4i}{\pi^{\rm 2}}\alpha^{\rm 3}+O(\alpha^{\rm 4}).
\label{32s}
\end{equation}

Now we evaluate the Fermi energy to the first order in $\alpha$. For
this purpose, it suffices to keep the first two terms in Eq.
(\ref{32s}) when substituting it into Eq. (\ref{12s}). Indeed, the
order of magnitude of the error $\delta \xi_{\rm i}$ does not
exceed $i_{\rm F}\alpha^{\rm 3}$ in this case. The error of
$\xi_{\rm i}^{\rm 2}$ is $2\xi_{\rm i}\delta \xi_{\rm i}$, and its
order of magnitude also does not exceed  $i_{\rm F}\alpha^{\rm
3}$, since $\xi_{\rm i}\leq 1$. The error of the sum $\sum_{\rm
i=1}^{\rm i_{\rm F}}\xi_{\rm i}^{\rm 2}$ is smaller than $i_{\rm
F}^{\rm 2}\alpha^{\rm 3}$, and the order of magnitude of the error
of the whole expression (\ref{12s}) does not exceed $i_{\rm
F}\alpha^{\rm 3}$. Since $ i_{\rm F}\simeq \sqrt{\varepsilon_{\rm
F}}/\alpha$, this estimation of the Fermi energy is correct to
first order in $\alpha$.

After the above-mentioned substitution into Eq. (\ref{12s}), we
have
\begin{equation}
\varepsilon_{\rm F}=\frac{\nu}{2\pi}\frac{1}{i_{\rm F}\alpha}+
\left(\frac{i_{\rm F}^{\rm 2}}{3}+\frac{i_{\rm
F}}{2}\right)\alpha^{\rm 2}- \frac{4i_{\rm F}^{\rm
2}}{3\pi}\alpha^{\rm 3}+O(\alpha^{\rm 2}). \label{33s}
\end{equation}
We divide the range of variation of  $\alpha$ into intervals
$(\alpha_{\rm i+1},\alpha_{\rm i})$, $i=2,3...$, so that
$\alpha_{\rm i}\equiv 1/l_{\rm i}$ we have $i_{\rm F}=i$ inside
these intervals. The values  $\alpha>0.3$, which correspond to
$l<l_{\rm min}$, are nonphysical. Let us find the boundaries of
the intervals.

From Eqs. (\ref{32s}), (\ref{33s}) and condition $\varepsilon_{\rm
F}=\xi_{\rm i+1}^{\rm 2}$ we can obtain an equation for
$\alpha_{\rm i+1}$:
\begin{equation}
\frac{8i^{\rm 3}}{3\pi}\alpha_{\rm i+1}^{\rm 4}-
\left(\frac{2i^{\rm 3}}{3}+\frac{3i^{\rm 2}}{2}\right)\alpha_{\rm
i+1}^{\rm 3}+ \frac{\nu}{2\pi}=0.                     \label{38s}
\end{equation}
According to Descartes' rule of signs, Eq. (\ref{38s}) has two
real positive roots. In zeroth approximation, one of them is
$\sim\sqrt{\varepsilon_{\rm F0}}/i$ and the other is of a higher
order of smallness. We are interested in the first root of Eq.
(\ref{38s}), because boundaries of the interval $(\alpha_{\rm
i+1},\alpha_{\rm i})$, with constant $i_{\rm F}$ inside, are
determined by this root. The boundaries are
\begin{equation}
\alpha_{\rm p}=\frac{1}{i}\sqrt{\varepsilon_{\rm
F0}}+\frac{1}{2i^{\rm 2}} \sqrt{\varepsilon_{\rm
F0}}\left(\frac{4}{\pi} \sqrt{\varepsilon_{\rm F0}}\mp 1\right).
\label{41s}
\end{equation}
The minus sign corresponds to  $p=i+1$, and the plus sign to
$p=i$. Thus, the width of the interval $(\alpha_{\rm
i+1},\alpha_{\rm i})$ decreases as $1/i^{\rm 2}$  with increasing
$i$ (or $l$), since $\alpha_{\rm i}-\alpha_{\rm
i+1}=\sqrt{\varepsilon_{\rm F0}}/i^{\rm 2}$.

Unexpectedly, the examination of function $\varepsilon_{\rm
F}(\alpha)$ indicates the insufficiency of its approximation by
the expression (\ref{33s}). This expression is stair-like
function, i.e., it is constant within intervals $(\alpha_{\rm
i+1},\alpha_{\rm i})$, where $i_{\rm F}=i$. It can be explained by
the fact that $\alpha_{\rm i+1}$ and $\alpha_{\rm i}$ differ in
the \emph{second} order of smallness, and after substituting into
(\ref{33s}) they give the same results (of course, terms of the
order more than \emph{first} must be omitted in the resultant
expression). A numerical calculation of $\varepsilon_{\rm F}$ by
the formula (\ref{33s}) leads to the error that is associated with
an incorrect consideration for terms of the order more than first
and gives an appearance of the oscillatory dependence
$\varepsilon_{\rm F}(l)$.

In order to take the terms $\sim \alpha^{2}$ on the right-hand
side of Eq. (\ref{33s}) into account, we must include the
following terms, which were earlier neglected:
\begin{equation}
\frac{1}{6}\alpha^{\rm 2}-\frac{2i_{\rm F}}{\pi}\alpha^{\rm 3}+
\frac{4i_{\rm F}^{\rm 2}}{\pi^{\rm 2}}\alpha^{\rm 4}. \label{44s}
\end{equation}
In this case, the  $\varepsilon_{\rm F}(l)$ dependence is
represented by a concave curve in each interval $(l_{\rm i},l_{\rm
i+1})$, $i=2,3,...$ ($l_{\rm i}=1/\alpha_{\rm i}$).

At the points $l=l_{\rm i}$, the derivative  $d\varepsilon_{\rm
F}/dl$ is discontinuous and its jump is $-2{\varepsilon_{\rm
F0}}^{\rm 3/2}/i^{\rm 2}$. To the left from this point, the
function grows, and to the right from it, the function decreases.
The jump in the derivative results in the appearance of cusps in
the plot. The sharpness of the cusps decreases with increasing
$i$.

For large values of $l$, Eq. (\ref{13s}) can be written as
\begin{equation}
i_{\rm F}=\frac{\sqrt{\varepsilon_{\rm F0}}}{\alpha}+O(\alpha^{\rm
0}). \label{51s}
\end{equation}

Substitute Eq. (\ref{51s}) into Eq. (\ref{33s}) and using
conventional units, we obtain
\begin{equation}
E_{\rm F}=E_{\rm F0}+\frac{\pi\hbar}{2}\sqrt{\frac{E_{\rm
F0}}{2m}}\left(1-\frac{8}{3\pi} \sqrt{\frac{E_{\rm F0}}{|U_{\rm
0}|}}\right)\frac{1}{L_{\rm z}}. \label{53s}
\end{equation}
The expression in the brackets is positive, i.e., asymptotically,
we always have  $E_{\rm F}>E_{\rm F0}$.

In this model, the work function is defined as:
\begin{equation}
W_{\rm e}=-U_{\rm 0}-E_{\rm F}, \label{W}
\end{equation}
and it is easy to see that $W_{\rm e}<W_{\rm e0}$. A role of the
size dependence of the bottom  of the potential well $U(L_{\rm
z})$ will be discussed later.

\subsection{Deformation force}

In order to calculate force characteristics, we must find the size
dependent electron kinetic energy. We denote the total kinetic
energy of the electrons by  $\varepsilon\equiv K/|U_{\rm 0}|$.

As noted above, the one-electron kinetic energy $\varepsilon_{\rm
p}$ is numerically equal to the square of the radius-vector of the
point in $\xi-$space. The contribution from the corresponding
element $dS$  of the disk to the total kinetic energy is
$d\varepsilon=\varepsilon_{\rm p}\sigma dS$, where the density of
states $\sigma$ is defined by Eq. (\ref{5s}). Next, we must
integrate over the disk area and sum the contributions from all
disks.

We introduce the notation  $\rho\equiv (\xi_{\rm xj}^{\rm
2}+\xi_{\rm ys}^{\rm 2})^{\rm 1/2}$. The maximum value of $\rho$
in the $i$-th disk is equal to the disk radius  $\rho_{\rm i}=
(\varepsilon_{\rm F}-\xi_{\rm i}^{\rm 2})^{\rm 1/2}$. We have
\begin{multline}
\varepsilon=4\pi l_{\rm x}l_{\rm y}\sum\limits_{\rm i=1}^{\rm
i_{\rm F}}\int\limits_{\rm 0}^{\rm \rho_{\rm i}}d\rho\rho
(\xi_{\rm i}^{\rm 2}+\rho^{\rm 2})\\= \pi l_{\rm x}l_{\rm
y}\left(i_{\rm F}\varepsilon_{\rm F}^{\rm 2}-\sum\limits_{\rm
i=1}^{\rm i_{\rm F}} \xi_{\rm i}^{\rm 4}\right). \label{56s}
\end{multline}
Performing the summation, we find
\begin{multline}
\varepsilon=\pi l_{\rm x}l_{\rm y}\left(\left(\frac{\nu^{\rm
2}}{4\pi^{\rm 2}}\frac{1}{i_{\rm F}\alpha^{\rm 2}}+
\frac{\nu}{3\pi}i_{\rm F}^{\rm 2}\alpha - \frac{4}{45}i_{\rm
F}^{\rm 5}\alpha^{\rm 4}\right)\right.\\+\left.
\left(\frac{\nu}{2\pi}i_{\rm F}\alpha - \frac{4\nu}{3\pi^{\rm
2}}i_{\rm F}^{\rm 2}\alpha^{\rm 2}- \frac{1}{6}i_{\rm F}^{\rm
4}\alpha^{\rm 4}+\frac{32}{45\pi}i_{\rm F}^{\rm 5}\alpha^{\rm
5}\right)\right)+O(\alpha). \label{59s}
\end{multline}

The asymptotic form of this expression in conventional units is
\begin{equation}
K=\frac{3}{5}N_{\rm e} E_{\rm F0}+\frac{3\pi\hbar}{8}N_{\rm e}
\sqrt{\frac{E_{\rm F0}}{2m}} \left(1-\frac{32}{15\pi}
\sqrt{\frac{E_{\rm F0}}{|U_{\rm 0}|}}\right)\frac{1}{L_{\rm z}}.
\label{61s}
\end{equation}
Let us discuss the origin of the different terms in Eq.
(\ref{61s}).

The first term in Eq. (\ref{61s}) is the kinetic energy in the
case, when the slab thickness $L_{\rm z}$  is comparable to other
dimensions. The distance between the disks in $\xi-$space in this
case is so small that the summation in Eq. (\ref{56s}) can be
replaced by integration.

The second term in the brackets in Eq. (\ref{61s}) appears for a
finite well depth. This correction is rather important and makes a
contribution of about 50\%. In contrast to the case of an infinite
well, electron localization in a well of finite depth is not
strict, therefore, the kinetic energy in the latter case is
smaller.

In order to compare our results with the experimental data
\cite{6.,7.}, we find the oscillating electron contribution to the
elastic force under the conditions of ideal plastic strains, i.e.,
in the case where total  volume of the slab is conserved:
$$
f_{\rm z}=-(\partial \varepsilon /\partial l_{\rm z})_{\rm V}.
$$
This part of the force has no relation to the phases of stretching
that are accompanied by a change in volume. It rather determines
the variation in slab elastic properties as the slab thickness is
varied. This force depends on the number of particles in the slab,
therefore, it is convenient to consider the force normalized by
$N_{\rm e}$,
\begin{multline}
\frac{F_{\rm z}}{N_{\rm e}}=\frac{\hbar^{\rm
2}}{2m}\left(-\frac{\pi \bar{n}_{\rm e}} {i}+\frac{2\pi^{\rm
2}}{3}\frac{i^{\rm 2}}{L_{\rm z}^{\rm 3}}-\frac{\pi^{\rm
3}}{9\bar{n}_{\rm e}}\frac{i^{\rm 5}}{L_{\rm z}^{\rm 6}} +\pi^{\rm
^{\rm 2}}\frac{i}{L_{\rm z}^{\rm 3}}\right. \\-
\left.\frac{4\pi^{\rm 2}}{k_{\rm 0}}\frac{i^{\rm 2}}{L_{\rm
z}^{\rm 4}}+\frac{5\pi^{\rm 3}}{24\bar{n}_{\rm e}} \frac{i^{\rm
4}}{L_{\rm z}^{\rm 6}}+\frac{16\pi^{\rm 3}}{15\bar{n}_{\rm
e}k_{\rm 0}} \frac{i^{\rm 5}}{L_{\rm z}^{\rm 7}}\right).
\label{67s}
\end{multline}

\section{CHARGED CLUSTERS}
\subsection{Two-component model: asymptotic expressions}

Let us consider now a neutral cluster, which contains $N_{\rm
e}/Z=N_{\rm i}=N$  atoms. Total energy of a cluster, charged by
$\Delta N_{\rm e} \ll N_{\rm e}$ electrons, can be written as
\cite{16.}:
\begin{equation}
\tilde{E}_{\rm N_{\rm e}+\Delta N_{\rm e}}= \tilde{E}_{\rm N_{\rm
e}}+\mu_{\rm e}\Delta N_{\rm e}+ \frac{(-e\Delta N_{\rm e})^{\rm
2}}{2C}, \label{000}
\end{equation}
where $\mu_{\rm e}$ is the electron chemical potential. A cluster
can retain the excess electrons $\Delta N_{\rm e}$ only when its
energy in this state is lower than the energy in the state with
$N_{\rm e}+\Delta N_{\rm e}-1$ electrons. By our definition, the
number of electrons in a cluster is critical, if the number of
electrons $\Delta N_{\rm e}^{\rm *}$, for which the reaction
$$
M_{\rm N_{\rm i}}^{\rm (\Delta N_{\rm e}^{\rm
*})-}\rightleftarrows M_{\rm N_{\rm i}}^{\rm (\Delta N_{\rm
e}^{\rm *}-1)-}+e^{\rm -}
$$
is reversible, and the ionization potential of the cluster,
$IP^{\rm *}>0$, tends to zero
\begin{equation}
\Delta \tilde{E}(\Delta N_{\rm e}^{\rm *})= \tilde{E}_{\rm N_{\rm
e}+\Delta N_{\rm e}^{\rm *}-1}-\tilde{E}_{\rm N_{\rm e}+\Delta
N_{\rm e}^{\rm *}}\equiv IP^{\rm *}
 \rightarrow 0.
\label{1.135}
\end{equation}
We note that the addition of one more surplus electrons to $\Delta
N_{\rm e}^{\rm *}$ is  possible only in a metastable state,
because the  affinity of this electron is
\begin{multline}
EA^{\rm *}  = \tilde{E}_{\rm N_{\rm e}+\Delta N_{\rm e}^{\rm *}} -
\tilde{E}_{\rm N_{\rm e}+\Delta N_{\rm e}^{\rm *}+1}  \\=
-\mu_{\rm e} - \frac{e^{\rm 2}}{2C}(2\Delta N_{\rm e}+1)<0.
\label{VAS-1}
\end{multline}
In any case, the relation
$$
IP^{\rm *}-EA^{\rm *}=\frac{e^{\rm 2}}{C}
$$
is valid. When  $\Delta N_{\rm e}>\Delta N_{\rm e}^{\rm *}$, the
cluster is overcharged. The surplus electrons are separated from
the free states by a potential barrier and they can be in the
metastable state for some time. The lifetime of each electron is
determined by specific conditions in the nonequilibrium system.

Using Eqs. (\ref{1.135}) and (\ref{000}),  we obtain for the
critical excess electron charge
\begin{equation}
\Delta N_{\rm e}^{\rm *}= \frac{W_{\rm e0}C-\mu_{\rm e1}}{e^{\rm
2}} +\frac{1}{2}, \label{1.136}
\end{equation}
where  $W_{\rm e0}=-\mu_{\rm e0}$ is the electron work function
for a flat surface, $\mu_{\rm e}=\mu_{\rm e0}+\mu_{\rm e1}/R$, and
$\mu_{\rm e1}/R$ is the first curvature correction term in the
electron chemical potential of a metal sphere of $R=N^{\rm
1/3}r_{\rm 0}$ radius with the notation $r_{\rm 0}$ for the mean
ion spacing.

Let us consider now a positively charged cluster of metal, which
contains $N_{\rm e}=ZN_{\rm i}$ electrons and  $N_{\rm i}+\Delta
N_{\rm i}$ ions. This situation is similar to that one when the
droplet with $N_{\rm i}$ ions ``contains''  $\Delta N_{\rm e}<0$
lacking electrons. In this connection, $\Delta N_{\rm e}$ has to
be divisible by $Z$.

The energy of charged cluster $\tilde{E}_{\rm N_{\rm i}+\Delta
N_{\rm i}}$ is related to the energy of a neutral cluster in the
following way
\begin{equation}
\tilde{E}_{\rm N_{\rm i}+\Delta N_{\rm i}}= \tilde{E}_{\rm N_{\rm
i}} +\mu_{\rm i}\Delta N_{\rm i}+\frac{(+eZ\Delta N_{\rm i})^{\rm
2}}{2C}. \label{1.137}
\end{equation}
As in Eq. (\ref{000}) the most essential size dependence  is due
to the self-repulsion of the surface charge $+eZ\Delta N_{\rm i}$.
Actually, the ions are not mobile and the redistribution in the
electronic subsystem mimics the distribution of an excess positive
charge.

The change in  the total energy, associated with the detachment of
$\Delta N_{\rm i}-$th ion, is
\begin{multline}
\Delta \tilde{E}(\Delta N_{\rm i})= \tilde{E}_{\rm N_{\rm
i}+\Delta N_{\rm i}-1} - \tilde{E}_{\rm N_{\rm i}+\Delta N_{\rm
i}}\\=-\mu_{\rm i}-\frac{e^{\rm 2}Z^{\rm 2}}{2C}(2\Delta N_{\rm
i}-1).
\end{multline}
A cluster with  charge  $+eZ\Delta N_{\rm i}$ can exist in
equilibrium only if  $\Delta \tilde{E}(\Delta N_{\rm i})>0$. So
the number of ions $\Delta N_{\rm i}^{\rm *}$ in a cluster is
critical, if the reaction
$$
M_{\rm N_{\rm i}+\Delta N_{\rm i}^{\rm *}}^{\rm Z(\Delta N_{\rm
i}^{\rm *})+} \rightleftarrows M_{\rm N_{\rm i}+\Delta N_{\rm
i}^{\rm *}-1}^{\rm Z(\Delta N_{\rm i}^{\rm *}-1)+}+M_{\rm 1}^{\rm
Z+}
$$
is reversible. In this case we have
\begin{equation}
\Delta N_{\rm i}^{\rm *} =\frac{W_{\rm i0}C-\mu_{\rm
i1}}{(Ze)^{\rm 2}} +\frac{1}{2}, \label{1.138}
\end{equation}
where  $W_{\rm i0}=-\mu_{\rm i0}$ is the ion work function for a
plane surface. For a sphere of radius  $R=(N_{\rm i}+\Delta N_{\rm
i})^{\rm 1/3}r_{\rm 0}$, using sum rules \cite{93}, we can write
\begin{equation}
\mu_{\rm i1}= \frac{2\gamma}{\bar{n}_{\rm e}}-\mu_{\rm
e1},\label{1.138bbb}
\end{equation}
where   $\gamma$ is the surface energy per unit area. For
materials under the study $\mu_{\rm e1}\simeq$ 1.9 eV$\times
a_{\rm 0}$ \cite{1996,94ab}.

If  $\Delta N_{\rm i}>\Delta N_{\rm i}^{\rm *}$, the cluster emits
a surplus ion, passing into the state with lower energy. This
approach corresponds to the consideration of the droplet as a
two-component electron-ion system.

The work function of single-charged ion can be expressed by means
of the Born cycle  \cite{168} using the ionization potential  of
single atom $IP(1)$, cohesive energy $\varepsilon_{\rm coh0}$, and
work function for a flat surface $W_{\rm e0}$:
\begin{equation}
W_{\rm i0}=\varepsilon_{\rm coh0}+IP(1)-W_{\rm e0}. \label{1.138a}
\end{equation}
For Pb  $\varepsilon_{\rm coh0}$ = 1.5 eV, $W_{\rm e0}$ = 4.0 eV,
$IP(1)$ = 7.4 eV, and we get  $W_{\rm i0}$ = 4.9 eV. When $R =
12a_{\rm 0}$, the critical charge is equal to +2.7$e$. This
estimation is in a good agreement with experimental data
\cite{Sattler} and with the results of more complicated
self-consistent calculations \cite{Nah-1}.

Our approximation assumes that the shape of a cluster is
unvariable during its charging. Eq. (\ref{1.136}) for $\Delta
N_{\rm e}^{\rm *}$ and Eq. (\ref{1.138}) for $\Delta N_{\rm
i}^{\rm *}$ take into account electron and ion emission and
distinguish between them. It is due to the necessity to expend the
energy on the embedding of a particle of this kind into the
cluster and on the redistribution of its charge over the surface.
Such a mechanism of the Coulomb instability  can be introduced as
an alternative to Rayleigh`s one. Estimates show that $\Delta
N_{\rm R}>\Delta N_{\rm i}^{\rm *}>\Delta N_{\rm e}^{\rm *}$, i.e.
during the charging, rather the single-electron/ion emission
occurs than Rayleigh's instability. For small clusters, the energy
quantization is important.

\subsection{Quantum spectra}

The shape of real charged clusters seldom looks like a sphere
(e.g. see \cite{15ao,dots}), therefore it is convenient to
determine the electron spectrum  using the parallelepiped model.
The allowable levels form a discrete spectrum. Wave vector
components are evaluated by solutions of Eq. (\ref{1s}) for each
directions. In order to separate real levels from virtual ones we
introduce the following condition:
\begin{equation}
k_{\rm p}/k_{\rm 0}<1. \label{VAS-0}
\end{equation}

Using the perturbation theory,  Eq. (\ref{1s}) can be reduced to
the infinite well problem \cite{KK-66}. Three relations of
identical form determine wave numbers, for example,
\begin{equation}
k_{\rm xj}=k_{\rm xj}^{\rm \infty}+\Delta k_{\rm xj}, \quad
\zeta\equiv |\Delta k_{\rm xj}/k_{\rm xj}^{\rm \infty}| \ll 1,
\label{KUBO}
\end{equation}
where  $k_{\rm xj}^{\rm \infty}=\pi j/L_{\rm x}$ is the solution
at $ k_{\rm 0}\rightarrow \infty$. Substituting Eqs. (\ref{KUBO})
into Eq. (\ref{1s}), we obtain in the first approximation for a
cube $\zeta=-2/k_{\rm 0}L$  and the energy spectrum
\begin{equation}
E_{\rm p}=\frac{\hbar^{\rm 2}\pi^{\rm 2}}{2mL^2}(j^{\rm 2}+s^{\rm
2}+i^{\rm 2})(1+ 2\zeta+O(\zeta^{\rm 2})). \label{KUBO0}
\end{equation}
One more alternative expression can be derived  from Eq. (2) under
the condition (44):

\begin{equation}
E_{\rm p} \simeq \frac{\hbar^{\rm 2}\pi^{\rm
2}}{2m}\left(\frac{k_{\rm 0}} {2+k_{\rm 0}L}\right)^{\rm 2}(j^{\rm
2}+s^{\rm 2}+i^{\rm 2}). \label{VAS-3}
\end{equation}

On the one hand, in a neutral cube number of electrons is given,
on the other hand it is determined by sum  2$\sum _{\rm p}
\delta(E-E_{\rm p})$ over all occupied levels taking into account
twofold spin degeneracy. Filling levels by electrons we find a
highest occupied state, $E^{\rm HO}<0$, counted off from the
vacuum level. Following the Koopmans` theorem the ionization
potential of a cubiform cluster can be determined as
\begin{equation}
IP=-E^{\rm HO}+\frac{e^{\rm 2}}{2C}. \label{KUBO1}
\end{equation}

\section{RESULTS AND DISCUSSION}

We performed calculation for slabs of trivalent Al, monovalent Au
and Na with electron concentration $\bar{n}_{\rm e}=3/4\pi r_{\rm
s}^{\rm 3}$, where  $r_{\rm s}=$ 2.07, 3.01  and 3.99 $a_{\rm 0}$,
respectively.  The work function values for semi-infinite metals,
which we used, are $W_{\rm e0}=$ 4.25, 5.15 (or 4.3)  and 2.7 eV
\cite{Mich,39}.

Figure \ref{Pg-Fig2} shows the results of calculation of the
thickness dependence of the work function for extended isolated
slabs. The inequality $W_{\rm e}<W_{\rm e0}$ is satisfied for the
whole thickness range. The values of the largest oscillations of
the work function are about 0.1 -- 0.2 eV. This size dependence is
in a general agreement with the experimental results \cite{14.}
and self-consistent calculations for extended thin Al slabs
\cite{12.} and cylindrical Al and Na wires \cite{11.,13.}.
However, there is a disagreement with the results of
\cite{8.,9.,10.,Woi}.
%%%%%%%%%%%%%%%%%%%%%%%%%%%%%%%%%%%%%%%%%%%%%%%%%%%%%%%%%%
\begin{figure}[!t!b!p]
\centering
\includegraphics [width=0.48\textwidth] {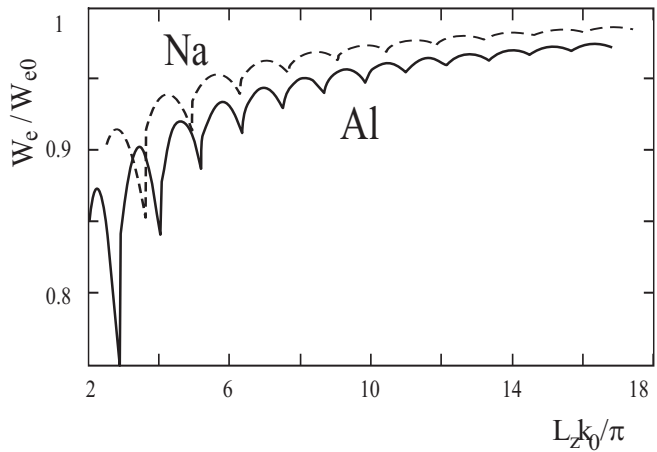}
\caption{Size dependence of the work function for metallic slabs.}
\label{Pg-Fig1}
\end{figure}
%%%%%%%%%%%%%%%%%%%%%%%%%%%%%%%%%%%%%%%%%%%%%%%%%%%%%%%%%%

Comparing $W_{\rm e}(L)$ for various metals, it is easily seen
that all differences are determined by the values of $r_{\rm s}$.
For aluminum (with the smallest $r_{\rm s}$), the amplitude of
oscillations of the work function  $W_{\rm e}/W_{\rm e0}$ is the
largest, the period  $\Delta L$  is the smallest, and position
$L_{\rm i}$ of the cusps are displaced to the left. These features
are well described by the approximate relations $1-W_{\rm
e}/W_{\rm e0}\sim 1/r_{\rm s}L$, $\Delta L \sim r_{\rm s}$, and
$L_{\rm i} \sim ir_{\rm s}$ ($i$ is the number of subband), which
follow from Eqs. (\ref{41s}) and (\ref{53s}).

In order to clarify the role of the thickness dependence of the
bottom of the well (see Eq. (\ref{Us})), we consider the data
presented in Table \ref{SH-1}. These data are extracted from the
results of the self-consistent calculations performed in
\cite{15.}. In that work, the electron energy spectrum was
calculated for a self-consistent spherical potential whose form
was far from being rectangular. In the rectangle-box model the
position of the first occupied level $E_{0}$ varies in accordance
with the position of flat bottom of the potential well. As the
well width is large (i.e., as $N\rightarrow\infty$), this level is
``lowered'' to the bottom. Hence, we can obtain quite reliable
information about the size dependence of the depth of the
rectangular well by determining the confinement behavior of the
lowest level in the potential profile corresponding to the
spherical cluster.
%%%%%%%%%%%%%%%%%%%%%%%%%%%%%%%%%%%%%%%%%%%%%%%%%%%%%%%%%%%%%%%%%%%%%%%%%%%
\begin{table}
\centering \caption{The energy of the first occupied state,
$E_{\rm 0}<0$, in spherical clusters Na$_{\rm N}$  calculated by
Ekardt \cite{15.}.} \vspace*{0.5cm}
%%%%%%%%%%%%%%%%%%%%%%%%%%%%%%%%%%%%%%%%%%%%%%%%%%%%%%%%%%%%%%%%%%%%%%%%%
\smallskip
\tabcolsep .1cm \label{SH-1}
\begin{tabular}{|c|ccccccc|} \hline\hline
N &18&20&34&40&58&68&90\\\hline $-E_{\rm 0}$ [eV]&5.10& 5.15&
5.41& 5.41& 5.59& 5.53 & 5.76\\
\hline \hline N &92&106&132&138&168&186&198\\ \hline
$-E_{\rm 0}$ [eV]&5.63& 5.64& 5.86&5.79& 5.91&5.92& 5.81\\
\hline \hline
\end{tabular}
\end{table}
%%%%%%%%%%%%%%%%%%%%%%%%%%%%%%%%%%%%%%%%%%%%%%%%%%%%%%%%%%%%%%%%%%

This dependence is almost monotonic and asymptotically weak.
Moreover, it does not compete with the thickness dependence of the
Fermi energy in Eq. (\ref{W}) and gives a minor contribution to
Eq. (\ref{53s}). Taking into account the dependence $U(L)$ in Eq.
(\ref{W}) leads to strengthening of the inequality $W_{\rm
e}<W_{\rm e0}$.

When the slab is inserted into contact with the electrodes, the
electron chemical potentials are equalized and the electronic
system should be considered as an open system with $W_{\rm
e}(L)=W_{\rm e0}$. The electrical neutrality of the slab cluster
is broken, and the part $\delta N_{\rm e}>0$ of the electron
liquid passes out to the bath. As a result, a contact potential
difference $\delta \phi$ appears.

In order to determine the contact potential difference, we
consider energy cycles, in which  electrons are transferred first
to infinity and then to the electrodes. By analogy with Eq.
(\ref{1.135}), we express the ionization potential of the slab,
having charge $+e\delta N_{\rm e}$, as
\begin{multline}
IP=E_{\rm N_{\rm e}-\delta N_{\rm e}-\Delta }-E_{\rm N_{\rm
e}-\delta N_{\rm e}}\\= W_{\rm e}\Delta+\frac{e^{\rm
2}}{2C}((\delta N_{\rm e}+\Delta )^{\rm 2}-\delta N_{\rm e}^{\rm
2})
\end{multline}
and write the electron affinity of the bath for the charge
$-e\Delta$  as $EA=W_{\rm e0}\Delta$. Equating these two
quantities, we obtain
\begin{equation}
W_{\rm e0}-W_{\rm e}-\frac{e^{\rm 2}}{2C}(2\delta N_{\rm e}+\Delta
)=0, \label{PC2}
\end{equation}
where C is the sample capacitance.

We note that  $\Delta $ can be infinitesimal, since an electron
can pass through the contact only partially (i.e., it can be
detected on both sides of the geometrical contact with a nonzero
probability).   $\delta N_{\rm e}$ can be considered as a smoothly
varying quantity. This situation is typical for one-electron
devices \cite{17.}.

We also assume that $C$ corresponds to the total capacitance of
the both contacts. The validity of this assumption depends on the
sample geometry and electromagnetic environment. This is not true
for a spherical cluster in contact with electrodes, but it is
valid for a cubiform cluster or a slab \cite{n1}. Neglecting
environment and setting $C=e\delta N_{\rm e}/\delta \phi $,
$\delta N_{\rm e}\ll N_{\rm e}$, and $\Delta \rightarrow 0$, we
obtain from Eq. (\ref{PC2})
\begin{equation}
\delta \phi =(W_{\rm e0}-W_{\rm e})/e. \label{PC3}
\end{equation}

Now the energy spectrum of the  $N_{\rm 1}=N_{\rm e}-\delta N_{\rm
e}$ electrons that remain in the slab must be found for the
rectangular potential well of  changed depth $U_{\rm 0}-e\delta
\phi $. The total kinetic energy $K_{\rm 1}$ of the remaining
electrons can be determined in the same way as for an isolated
slab but with changed energy spectrum and  number of electrons.
For the oscillating part of the elastic force, we have $F_{\rm
z1}=-\left(\partial\Omega /\partial L_{\rm z}\right)_{\rm V}$,
where $\Omega =K_{\rm 1}+W_{\rm e0}N_{\rm 1}$ is the thermodynamic
potential.

Using the data from Fig. 1, we can also determine the contact
potential difference  $\delta \phi$. This potential difference
produces a negative shift of the well depth (for the thinnest
slab, it attains 0.5 -- 1 eV). This leads to a displacement of
density of states to values corresponding to greater thicknesses.
The stretched sample acts as an ``electron pump''  with respect to
the electrodes, alternately ejecting electronic liquid and drawing
it back.

In Fig. \ref{Pg-Fig2}, we show the part of the elastic force that
is due to the quantization. As can be seen from this figure, the
amplitude of the force oscillations depends strongly on $r_{\rm
s}$. For sodium, the oscillation amplitude is 8 times smaller than
that for Aluminum. The character of this dependence can be
determined from Eq. (\ref{67s}): $(F_{\rm z}/N_{\rm e})_{\rm i}
\sim ir_{\rm s}^{\rm 3}$. Our estimations showed that, for a slab
in a contact, force oscillations are analogous in shape and
amplitude.
%%%%%%%%%%%%%%%%%%%%%%%%%%%%%%%%%%%%%%%%%%%%%%%%%%%%%%%%%%
\begin{figure}[!t!b!p]
\centering
\includegraphics [width=0.48\textwidth] {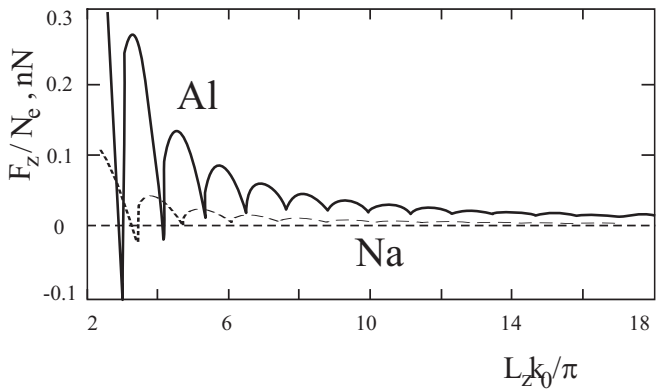}
\caption{Size dependence of the oscillating component of elastic
force $F_{\rm z}/N_{\rm e}$ for metallic slabs.} \label{Pg-Fig2}
\end{figure}
%%%%%%%%%%%%%%%%%%%%%%%%%%%%%%%%%%%%%%%%%%%%%%%%%%%%%%%%%%

The first maximum of the oscillating part of the force  $F_{\rm
z}/N_{\rm e}$ for Au (corresponding to a slab thickness of one
monolayer) is 0.2 nN, i.e., it is much smaller than the
experimental value for a wire, which is equal to 1.5 nN \cite{2.}.
This difference can be explained by both the difference in the
dimensionality of the electron gas in a wire and in a slab and the
effect of current flowing through a contact estimated in Ref.
\cite{20.} and discussed in Ref. \cite{Vetra}. Note that there are
no experimental results for slabs.

\begin{figure}[!t!b!p]
\centering
\includegraphics [width=0.48\textwidth] {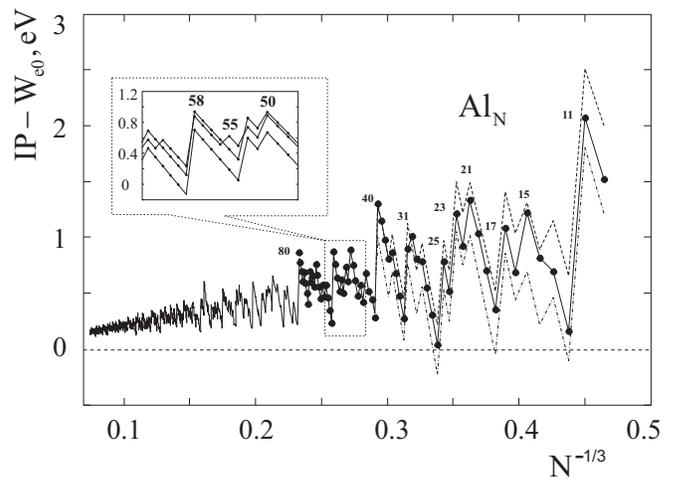}
\caption{Size dependences of the first ionization potential
(\ref{KUBO1}) of Al$_{\rm N}$ clusters. Solid, dashed, and
dot-dashed lines indicate the dependences determined according to
solutions Eqs. (\ref{1s}), (\ref{KUBO0}), and (\ref{VAS-3}),
respectively. Numbers at the top correspond to the numbers of
atoms in the cluster.} \label{Pg-Fig3}
\end{figure}

For the case of aluminum cluster, let us apply analytical approach
of the previous section. Size dependences of the ionization
potential (\ref{KUBO1}) calculated for the cubiform clusters
Al$_{\rm N}$ are shown in Fig. \ref{Pg-Fig3}. Here we use spectra,
which are determined by Eqs. (\ref{1s}), (\ref{KUBO0}), and
(\ref{VAS-3}). For the  $N$ range  (10, 3000), calculations
designate an essential role of spectrum quantization even for
large clusters. Magic numbers obtained are close to those found
experimentally \cite{38}. They are different in the case of single
valence sodium clusters.
%%%%%%%%%%%%%%%%%%%%%%%%%%%%%%%%%%%%%%%%%%%%%%%%%%%%%%%%%%
%%%%%%%%%%%%%%%%%%%%%%%%%%%%%%%%%%%%%%%%%%%%%%%%%%%%%%%%%%

Beginning from one hundred atoms, calculations of the spectrum
based on approximate  expressions (\ref{KUBO0}) and (\ref{VAS-3})
give quite reasonable results. However, their inaccuracy results
in the level hierarchy  differing from that determined from Eq.
(\ref{1s}). There is a difference  between spectra calculated by
Eqs. (\ref{KUBO0}) and (\ref{VAS-3}) for $N$ values near $N = 58$
(see insert in Fig. 3) \cite{n2}.

At the next step we investigate the ionization potential of a
cluster as a function of its shape. We assume that  shape of a
cuboid cluster varies from strongly flattened to elongated. Thus,
we have a  monatomic slab of thickness $L$ at the beginning and  a
monatomic chain of length $L$ at the end. During this evolution,
the sample volume is supposed to be constant and equal to 4
nm$^{\rm 3}$.

We divide the range of the size variation  into 10$^{\rm 3}$
intervals and find the spectrum for each of them using Eq.
(\ref{1s}). We exchange a capacitance of the parallelepiped by
capacitance of equivalent spheroid in the ionization potential
(\ref{KUBO1}). The size dependence of the capacitance has a
minimum for a sphere. In the limiting cases of the slab and the
wire of monoatomic thickness  capacitances are approximately twice
and sevenfold larger, respectively.

\begin{figure}[!t!b!p]
\centering
\includegraphics [width=0.48\textwidth] {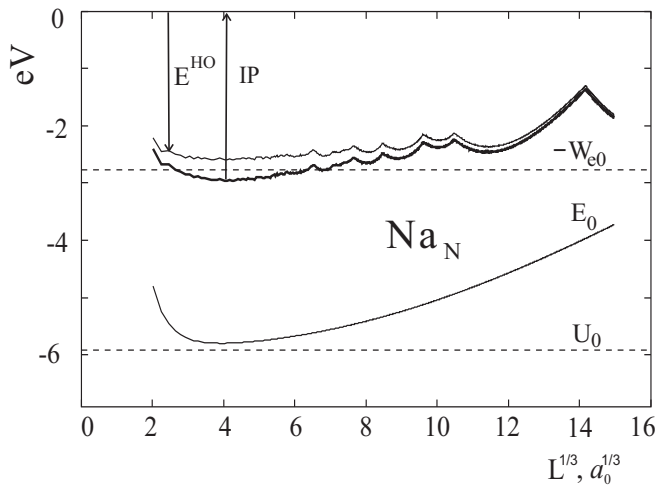}
\caption{Evolution in the size dependence of the first ionization
potential (\ref{KUBO1}) of  Na$_{\rm N}$ cubiform cluster with a
change in the cluster shape from a slab to a wire. Dashed lines
represent the bottom position of the potential well, $U_{\rm 0}$,
and work function for the flat surface, $W_{\rm e0}$.}
\label{Pg-Fig4}
\end{figure}

Figure \ref{Pg-Fig4} displays the behavior of the electron work
function and the ionization potential of the isolated sodium
samples of varying shapes. The inequality  $-E^{\rm HO} < W_{\rm
e0}$  is observed to be obeyed over the whole range of the
considered dimensionalities. The size dependence of energy of the
first occupied state, $E_{\rm 0}(L)$, has minimum at the point
corresponding to a cubiform cluster. As it can be seen, there are
ranges of lengths, where $IP<W_{\rm e0}$ and $IP>W_{\rm e0}$. The
inequality $IP<W_{\rm e0}$ is rather surprising. It is known from
experiments that the work function $W_{\rm e0}$ of alkali metals
is approximately equal to the one half of $IP$  of the atom
\cite{79a}. Therefore one would expect that the value of $IP$ of a
small solid sample belongs to the interval $W_{\rm e0}< IP<
IP({\rm atom})$ independently of the shape of sample surface.
However, the competition between the size correction $W_{\rm
e}(L)$  and the $e^{\rm 2}/2C$ term in expression (\ref{KUBO1})
for the ionization potential can lead to the opposite inequality.
%%%%%%%%%%%%%%%%%%%%%%%%%%%%%%%%%%%%%%%%%%%%%%%%%%%%%%%%%%
%%%%%%%%%%%%%%%%%%%%%%%%%%%%%%%%%%%%%%%%%%%%%%%%%%%%%%%%%%

Using mass-spectrometer \cite{Yannou}, minimal number of atoms $N$
have been recently determined for which an existence of stable
charged clusters  Au$_{\rm N}^{\rm 2-}$ ($N>$27), Au$_{\rm N}^{\rm
3-}$ ($N>$58), and  Ag$_{\rm N}^{\rm 2-}$ ($N>$27) is possible
under the condition of a  particle retaining 2 or 3 surplus
electrons. This problem is inverse to the one considered above.
Here $\Delta N_{\rm e}^{\rm *}$ is a parameter and $N$ is unknown.

We note that even for particles containing more than a thousand of
ions the critical charge does not exceed a few units. This
interesting feature is a result of the strong Coulomb repulsion of
the surplus charge spread over the surface of the particle. The
situation is different for atomic and molecular ions, where
electrons are not collectivized.

For calculations, we use the following experimental values of
surface tension: $\tau$ = 1134, 780, 191 erg/cm$^{\rm 2}$ for Au,
Ag, and Na, respectively. Here, we assume that the surface energy
equals the surface tension, while in reality these values can be
considerably different from \cite{Kurb}.

Rayleigh`s expression gives  numbers of atoms in the critical
clusters, which are 4 -- 5 times less. These numbers are $N\approx
9$ and 6, for Au$_{\rm N}^{\rm 3-}$ and Ag$_{\rm N}^{\rm 2-}$,
respectively. In our approach the problem is reduced to the
solution of the equation
\begin{multline}
IP^{\rm *}(\Delta N_{\rm e}^{\rm *},N)=-E^{\rm HO}(\Delta N_{\rm
e}^{\rm *},N)\\-\frac{e^{\rm 2}}{2C_{\rm \rm eff}(N)}(2\Delta
N_{\rm e}^{\rm *}-1)=0. \label{KUBO2}
\end{multline}

Effective capacitance $C_{\rm \rm eff}=R+\delta$ is used in order
to explain experimental results for charged clusters. The
additional small quantity $\delta$ is caused by an increase of
radius of the charging electron ``cloud''. The $\delta$ value was
introduced for calculations of the polarization by Snider and
Sorbello \cite{42} and the ionization potential of clusters by
Perdew \cite{211.}. The averaged dependence $\delta(r_{\rm 0})
=1.617+0.199(r_{\rm 0}/Z^{\rm 1/3}-2.07)$ (in bohrs) is obtained
using coordinates of the image plane for various crystallographic
faces \cite{1996}, which were computed in the stabilized jellium
model \cite{KiejnaW}.

Note that introduction of $\delta$  to Eqs. (\ref{KUBO1}) and
(\ref{KUBO2}) is not rigorous. This procedure accounts only for
the Hartree contribution, $\delta /R^{2}$, into correction term
$\sim 1/R^{2}$ in the energy $1/R-$expansion. Nevertheless,
solution of Eq. (\ref{KUBO2}) is responsible to the $\delta$ value
\cite{n3}.  With discussed modification for $IP$, Eq.
(\ref{1.135}) is adequate for explaining consecutive
fotoionization acts for large clusters Al$_{N}$  in the wide $N$
range  (2000, 32000) \cite{Hoffmann}. The use of $C_{\rm
eff}=R+\delta$  makes the size monotonic component of $IP(N)$ in
Fig. 3 some weaker.

\begin{figure}[!t!b!p]
\centering
\includegraphics [width=0.48\textwidth] {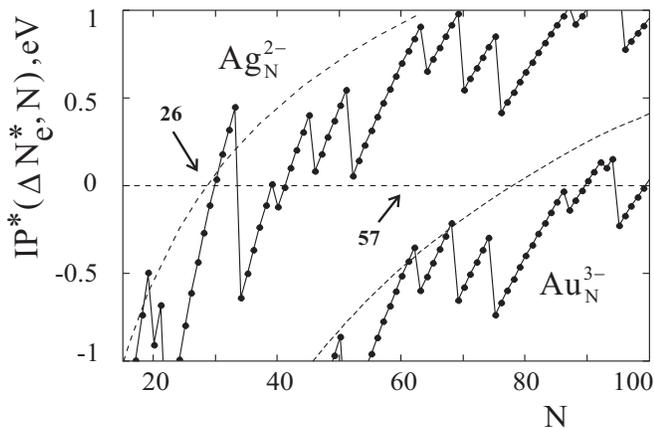}
\caption{Size dependences of the ionization potential
(\ref{KUBO2}) Au$_{\rm N}^{3-}$ and Ag$_{\rm N}^{2-}$ clusters
(solid line). Dashed lines represent the quasiclassical
dependences  $IP(N)$ (\ref{1.135}). Arrows indicate the
experimental critical numbers $N^{*}$.} \label{Pg-Fig5}
\end{figure}

The size dependence $IP^{*}(\Delta N_{\rm \rm e}^{*},N)$
calculated from relationships (\ref{KUBO2}) and (\ref{1.136}) is
shown in Fig. \ref{Pg-Fig5}. Points of the diagrams with
$IP^{*}=0$ indicate $N$ values for the critical clusters. It can
be seen from Fig. 5 that the quasiclassical dependence
(\ref{1.136}) and Eq. (\ref{KUBO2}) with inclusion of the level
quantization leads both to the better agreement with  experimental
data than Rayleigh`s formula. For Au $_{\rm N}^{3-}$, we use the
value $W_{\rm e0}$ = 5.15 eV recommended by Michaelson
\cite{Mich}. The Au$_{\rm N}^{2-}$ clusters appear to be stable
when $N>20$. However, with another value $W_{\rm e0}$ = 4.3 eV,
proposed by Fomenko \cite{39}, Eq. (\ref{KUBO2}) gives Au$_{\rm
27}^{2-}$ and Au$_{\rm 110}^{3-}$ critical clusters. Specific
features in the energy properties of gold clusters were noted by
Garron \cite{Garron}.
%%%%%%%%%%%%%%%%%%%%%%%%%%%%%%%%%%%%%%%%%%%%%%%%%%%%%%%%%%
%%%%%%%%%%%%%%%%%%%%%%%%%%%%%%%%%%%%%%%%%%%%%%%%%%%%%%%%%%

Finally, we apply our computation procedure to the case of
positively charged clusters Na$_{\rm N}^{n+}$. N\"{a}her \emph{et
al.} \cite{Nah-2}  determined experimentally critical numbers $N$
= 64, 123, and 208 for clusters with $n$ = 3, 4, and 5,
respectively. In our model, we have $N\equiv N_{\rm i}+\Delta
N_{\rm i}^{*}$ ,  and $n\equiv\Delta N_{\rm i}^{*}$. The critical
sizes are calculated using Eqs. (\ref{1.138}) -- (\ref{1.138a}) in
which we replace $W_{\rm e0}\rightarrow -E^{\rm HO}$ and eliminate
$\mu_{\rm e1}$. We use $\varepsilon_{\rm coh}$ = 1.13 eV and
$IP(1)$ = 5.14 eV in the calculations.

\begin{figure}[!t!b!p]
\centering
\includegraphics [width=0.48\textwidth] {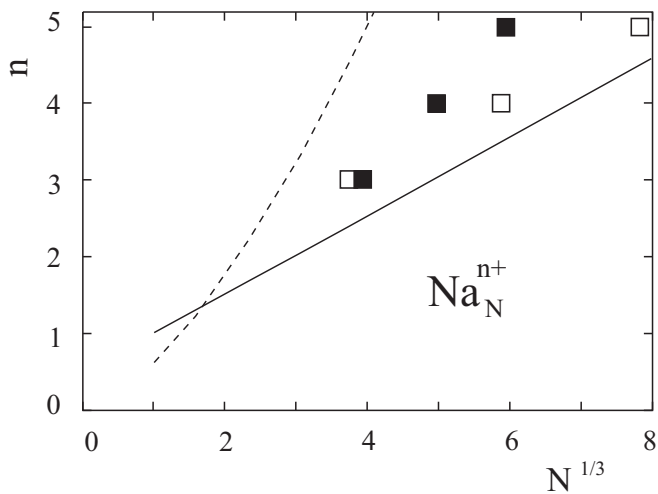}
\caption{Critical sizes of positively charged clusters Na$_{\rm
N}^{n+}$. Solid line -- quasiclassical dependence (\ref{1.138}),
dashed line -- the Rayleigh dependence, ($\blacksquare$) --
experiments, and ($\square$) are the results of quantization.}
\label{Pg-Fig6}
\end{figure}

Our results for the Na$_{\rm N}^{n+}$ clusters are presented in
Fig. \ref{Pg-Fig6}. For the small sized clusters, the situation
%%%%%%%%%%%%%%%%%%%%%%%%%%%%%%%%%%%%%%%%%%%%%%%%%%%%%%%%%%
%%%%%%%%%%%%%%%%%%%%%%%%%%%%%%%%%%%%%%%%%%%%%%%%%%%%%%%%%%
can be described by the Rayleigh`s formula according to which
$|Q_{\rm R}|\propto N^{1/2}$. The quasiclassical instability leads
to the relationship $eZ\Delta N_{\rm i}^{*}\propto N^{1/3}$.
Calculated critical sizes of the clusters occur to be overrated as
compared to experimental values. From the data presented in Fig.
5, one can assume that  cubiform clusters transforms predominantly
into cuboids. The change in energy of the highest occupied state
$E^{\rm HO}$ is significantly less than that associated with the
charging due to an increase in capacitance.  Elongation of
clusters is accompanied also by the change in the cohesion energy
(see Eqs. (\ref{1.138}) and (\ref{1.138a})). This is confirmed by
experimental data on deformation of point contacts. A decrease in
the dimensionality gives a considerable increase in strength of a
contact \cite{PR-2003}. These factors can be responsible for the
difference between dependences $\Delta N_{\rm i}^{*}(N)$
calculated and experimentally found.

\section{Summary}

The problem of energetics of finite metallic systems has been
considered. We have developed an analytical theory of size
dependent energy and force characteristics for metal slabs by
using rectangular-box model, free-electron approximation and the
model of a potential well of finite depth. This approach makes it
possible to calculate the oscillatory size dependence of the work
function. It is concluded that the work function of
low-dimensional metal structure is always smaller that that of the
semi-infinite metal. The theory has been applied to calculate the
elastic effects in thin metallic slab.

In the framework of the two-component model, the size dependent
Coulomb instability of charged metallic clusters has been
described. A mechanism of the Coulomb instability in charged
metallic clusters, different from Rayleigh's one, has been
discussed. We showed that the two-component model of a metallic
cluster in quasiclassical approximation leads  to different
critical sizes depending on a kind of charging particles
(electrons or ions). For the cubiform clusters, the electronic
spectrum quantization has been taken into account. The model
enables us to discover the physical origin of the instability and
to explain the critical sizes of charged Au$_{\rm N}^{\rm n-}$,
Ag$_{\rm N}^{\rm n-}$ and Na$_{\rm N}^{\rm n+}$ clusters of
different shape. Results of this investigation might have an
application in diagnostics of ultradispersed media,
single-electronics, and in nanotechnology.

\acknowledgments{We are grateful to Dr. W. V. Pogosov for reading
the manuscript. This work was supported by the Ministry of
Education and Science of Ukraine (Programme ``Nanostructures''). }

%%%%%%%%%%%%%%%%%%%%%%%%%%%%%%%%%%%%%%%%%%%%%%%%%


\begin{thebibliography}{0}
\expandafter\ifx\csname natexlab\endcsname\relax\def\natexlab#1{#1}\fi
\expandafter\ifx\csname bibnamefont\endcsname\relax
  \def\bibnamefont#1{#1}\fi
\expandafter\ifx\csname bibfnamefont\endcsname\relax
  \def\bibfnamefont#1{#1}\fi
\expandafter\ifx\csname citenamefont\endcsname\relax
  \def\citenamefont#1{#1}\fi
\expandafter\ifx\csname url\endcsname\relax
  \def\url#1{\texttt{#1}}\fi
\expandafter\ifx\csname urlprefix\endcsname\relax\def\urlprefix{URL }\fi
\providecommand{\bibinfo}[2]{#2}
\providecommand{\eprint}[2][]{\url{#2}}

\end{thebibliography}


\begin{references}

\bibitem{14.}  J. J. Paggel, C. M. Wei, M.Y. Chou, D.-A. Luh, T. Miller,
and T.-C. Chiang, Phys. Rev. B \textbf{66} 233403 (2002).

\bibitem{Otero} R. Otero,  A. L. Vazquez de Parga, R. Miranda,
 Phys. Rev. B \textbf{66} 115401 (2002).

\bibitem{Chulok}  E. Ogando, N. Zabala, E. V. Chulkov, and M.J. Puska,
Phys. Rev. B \textbf{69} 153410 (2004).

\bibitem{6.}  C. Untiedt, G.~Rubio, S.~Vieira, and N.~Agra\"{\i }t,
Phys. Rev. B \textbf{56} 2154 (1997).

\bibitem{7.}  G. Rubio-Bollinger, S. R. Bahn, N. Agra\"{\i }t,
K. W. Jacobsen, and S. Vieira, Phys. Rev. Lett. \textbf{87} 026101
(2001).

\bibitem{1.} J.P. Rogers III, P. H. Cutler, T. E. Feuchtwang,
 and A.A. Lucas, Surf. Sci. \textbf{181} 436 (1987).

\bibitem{Nag} E. L. Nagaev, Phys. Rep. \textbf{222} 201 (1992).

\bibitem{2.}  M. V. Moskalets,  Pis'ma v Zh. Exp. Teor.
Fiz. \textbf{62} 702 (1995) [JETF Lett. \textbf{62} 719 (1995)].

\bibitem{3.}  J. M.~van Ruitenbeek, M. H.~Devoret, D.~Esteve,  and C.
Urbina, Phys. Rev. B \textbf{56} 12566 (1997).

\bibitem{4.}  C. A.~Stafford, D.~Baeriswyl,  and J. B\"{u}rki, Phys. Rev.
Lett.  \textbf{79} 2863 (1997).

\bibitem{5.}  S.~Blom, H.~Olin, J. L.~Costa-Kramer, N.~Garcia, M.
Jonson, P. A.~Serena,  and R. I.~Shekhter, Phys. Rev. B
\textbf{57} 8830 (1998).

\bibitem{PR-2003}N. Agra\"{\i}t, A.L. Yeyati,  and J. M. van
Ruitenbeek, Phys. Rep.  \textbf{377} 81 (2003).

\bibitem{Schulte}   F. K. Schulte, Surf. Sci. \textbf{55} 427 (1976).

\bibitem{Pash}  A. M. Gabovich, L. G. Ilchenko,  and E. A. Pashitskii,
 Fiz. Tverd. Tela  \textbf{21} 1683 (1979) (in Russian).

\bibitem{8.}P. J. Feibelman,  and D. R. Hamann, Phys. Rev. B \textbf{29}
6463 (1984).

\bibitem{9.}J. C. Boettger, Phys. Rev. B \textbf{53} 13133 (1996).

\bibitem{10.}A. Kiejna, J. Peisert,  and P. Scharoch, Surf. Sci. \textbf{54}
432 (1999).

\bibitem{Woi}K.F. Wojciechowski, Phys. Rev. B \textbf{60} 9202 (1999).

\bibitem{11.}N. Zabala, M. J. Puska,  and R. M. Nieminen, Phys. Rev. B
\textbf{59} 12652 (1999).

\bibitem{12.}  I.~Sarria, C.~Henriques, C.~Fiolhais,  and J. M.~Pitarke,
Phys. Rev. B \textbf{62} 1699 (2000).

\bibitem{13.}E. Ogano, N. Zabala,  and M. J. Puska, Nanotechnology
\textbf{13} 363 (2002).

\bibitem{Sattler} K. Sattler, O. M\"{u}hlbach,  and E. Recknagel, Phys. Rev.
Lett. \textbf{47} 160 (1981).

\bibitem{Nah-1} U. N\"{a}her, S. Bjornholm, S. Frauendorf, F. Garcias,
 and C. Guet, Phys. Rep. \textbf{285} 245 (1997).

\bibitem{Nah-2}U. N\"{a}her, H. G\"{o}hlich, T. Lange,  and T. P. Martin,
Phys. Rev. Lett. \textbf{68} 3416 (1992).

\bibitem{Yannou}C. Yannouleas, U. Landman, A. Herlert,  and L.
Schweikhard, Phys. Rev. Lett. \textbf{86} 2996 (2001).

\bibitem{Duft}D. Duft, H. Lebius, B. A. Huber, C. Guet,  and T.
Leisner, Phys. Rev. Lett. \textbf{89} 084503 (2002).

\bibitem{93} I. T. Iakubov, A. G. Khrapak, V. V. Pogosov,  and S.A.
Trigger,  Phys. Stat. Sol. b \textbf{145} 455 (1988).

\bibitem{1996} A. Kiejna  and V. V. Pogosov, J. Phys.: Cond.
Matter \textbf{8} 4245 (1996).

\bibitem{Mich}H. B. Michaelson,  J. Appl. Phys. \textbf{48} 4729 (1977).

\bibitem{16.}I. T.~Iakubov, A. G.~Khrapak, L. I.~Podlubny, and
V. V.~Pogosov, Solid State Commun. \textbf{53} 427 (1985).

\bibitem{94ab} M. Seidl, J. P. Perdew, M. Brajczewska,  and C. Fiolhais,
Phys. Rev. B \textbf{55} 13288 (1997); J. Chem. Phys. \textbf{108}
8182 (1998).

\bibitem{168} F. Vericat  and M. P. Tosi,  Nuovo Cimento D \textbf{8}
105 (1986).

\bibitem{15ao} D. M. Wood and  N. W. Ashcroft, Phys. Rev.
 B \textbf{25} 6255 (1982).

\bibitem{dots}S. M. Reimann and  M. Manninen, Rev. Mod. Phys.
\textbf{74} 1283 (2002).

\bibitem{KK-66}A. Kawabata and  R. Kubo, J. Soc. Jap.\textbf{ 21} 17 (1966).

\bibitem{39}V. S. Fomenko, \emph{Emission Properties of
Materials} (Naukova Dumka, Kiev, 1981).

\bibitem{15.}W. Ekardt, Phys. Rev. B \textbf{29} 1558 (1984).

\bibitem{17.} D. M. Kaplan, V. A. Sverdlov,  and K. K. Likharev,
Phys. Rev. B \textbf{68} 045321 (2003).

\bibitem{n1}Near the edges of the parallelepiped, the surface
distribution of excess positive charge should be similar to that
for real ionization.

\bibitem{20.}M. Brandbyge, J.-L. Mozos, P. Ordejon, J. Taylor,  and K. Stokbro,
Phys. Rev. B \textbf{65} 165401 (2002).

 \bibitem{Vetra} M. Di Ventra, Y.-C. Chen,  and T. N. Todorov, Phys. Rev.
 Lett. \textbf{92} 176803 (2004).

\bibitem{38} W. A. de Heer, Rev. Mod. Phys. \textbf{65} 611 (1993).

\bibitem{n2} The procedure of computation of spectrum  has some specific
feature. We obtain the spectral term $E^{\rm HO}$ for a cuboid
combining of solutions of the one-dimensional problem (see, e.g.
Eq. (\ref{KUBO0})), therefore it becomes necessary to choose the
combination that realizes the minimum energy.

\bibitem{79a}K. Wong, S. Vongehr,  and V. V. Kresin, Phys. Rev. B
\textbf{67} 035406 (2003).

\bibitem{Kurb} A. Kiejna and V. V. Pogosov, Phys. Rev.
B \textbf{62} 10445 (2000); V. V. Pogosov and  V.P. Kurbatsky, Zh.
Exper. Teor. Fiz. \textbf{119} 350 (2001)  [JETF. \textbf{92} 304
(2001)]; V. V. Pogosov\textbf{,} O. M. Shtepa  (cond-mat/0310176).

\bibitem{42} D. R. Snider and  R. S. Sorbello, Phys. Rev. B \textbf{28}
5702 (1983).

\bibitem{211.}J. P. Perdew,  Phys. Rev. B \textbf{37} 6175 (1988).

\bibitem{KiejnaW}A. Kiejna,  K.F. Wojciechowski,
\emph{Metal Surface Electron Physics} (Pergamon, Oxford, 1996).

\bibitem{n3}However, this solution is not sensitive to the
self-compression effect \cite{1996}.

\bibitem{Hoffmann}M. A. Hoffmann, G. Wrigge , B. von Issendorff,
Phys. Rev. B \textbf{66} 041404 (2002).

\bibitem{Garron}R. Garron,  Ann. Phys. \textbf{10} 595 (1965).
\end{references}
\end{document}